# Determination of the local contrast of interference fringe patterns using continuous wavelet transform

Jong Kwang Hyok , Kim Chol Su

Institute of Optics, Department of Physics, **Kim Il Sung** University, Pyongyang, DPR of Korea

**Abstract**

The contrast of fringe patterns cannot be uniform over the whole interferogram because of various causes such as non-uniformity of reflectance, air fluctuation or vibration, and sometimes it is necessary to analyze it. In this paper we propose a method of determining the local contrast of interference fringe patterns using continuous wavelet transform.

## 1. Introduction

In optical interferometry the physical quantities to be considered modulate the intensity of the fringe patterns as follows.

$$I(x, y) = I_0 \cdot [1 + \gamma(x, y) \cdot \cos \phi(x, y)] , \qquad (1)$$

where $I(x, y)$ represents the intensity of the pattern image, $I_0$ is the background intensity, $\gamma(x, y)$ is the local contrast, and $\phi$ is the optical phase related to the physical quantities to be considered. In interferometry phase retrieval from the fringe pattern is therefore essential if one wants to recover the physical quantities to be considered.

To recover the phase information from a single interferogram requires that it should be modulated with a suitable spatial carrier frequency. But as shown in Eq. (1), the fringe quality is strongly influenced by local contrast $\gamma(x, y)$ and it is important to determine it to enhance the accuracy of fringe analysis. For these type of fringe pattern there are three well known methods for phase recovery and fringe contrast, namely

Fourier transform method[1], the windowed Fourier transform[2,3] and wavelet transform[4]. The Fourier transform method is based on the filtering of Fourier spectrum of the fringe pattern. But in the case of rapid variation of spatial frequency or direction of fringe pattern, it is difficult to filter the effective spectrum accurately only by Fourier transform. Furthermore, in Fourier transform method it is impossible to verify the corresponding relation between local intensities and local fringe frequencies. Therefore it is difficult to separate DC component and noise spectrum from effective spectrum and to analyze a local area of fringe pattern. In the windowed Fourier transform, it is possible to analyze the local character of the fringe pattern. But in the case of rapid variation of spatial frequency, the accuracy of fringe analysis is not so high because of the fixed size of window.

Modulated fringe patterns can be best described as non-stationary signals with local instantaneous frequencies that are proportional to the gradient of the phase distribution of the measured field. The continuous wavelet transformation, which is a correlation of a signal with a "mother" wavelet with a certain scale and shift, can be used to verify the correspondence between the local fringe intensity and contrast.

## 2. Determination of the local contrast of interference fringe patterns

CWT is represented by the correlation between the given signal $f$ and a wavelet functions $\psi$ [5].

$$W_\psi f(a,b) = \frac{1}{\sqrt{a}} \int_{-\infty}^{+\infty} f(x)\psi\left(\frac{x-b}{a}\right)dx ,  \qquad (2)$$

where $W_\psi f(a,b)$ is the coefficient matrix of CWT, $f(x)$ is the signal, and $\psi\left(\frac{x-b}{a}\right)$ is the wavelet function with scale parameter $a$ and shift parameter $b$.

The scale parameter describes the feature of fringe, such as fringe frequencies, and the shift parameter specifies a local region where the signal is analyzed. The CWT transforms the signals in the spatial domain into the space-frequency domain. The local signal analysis approach is a major advantage of CWT compared a Fourier transform.

Because the CWT is essentially a correlation operation between a signal and a wavelet, the analysis accuracy of the fringe pattern greatly depends on the choice of the mother wavelet function. As shown in Eq. (1), the interference fringe pattern is cosine modulated, so a suitable choice of wavelet function is the complex Morlet wavelet.

$$\psi(x) = \frac{1}{\sqrt{2\pi}} \exp\left(-\frac{x^2}{2}\right) \exp(i\omega_0 x) ,  \qquad (3)$$

where $\omega_0$ is the frequency of the Morlet wavelet. The terms $\exp(-x^2/2)$ and $\exp(i\omega_0 x)$ represent a fast-decaying Guassian envelope and a complex function with sinusoidal characteristics respectively.

Eq. (1) can be rewritten in terms of the fringe frequencies $\omega_s$ as follows

$$I(x, y) = I_0 \cdot [1 + \gamma(x, y) \cdot \cos(\varpi_s x)] \qquad (4)$$

The wavelet transformation of Eq. (4) using the complex Morlet wavelet is as follows.

$$W_\psi f(a,b) = \sqrt{2\pi} I_0 \exp\left(-\frac{\omega_0^2}{2}\right) + \frac{\sqrt{a}}{2} I_0 \gamma \exp\left(-\frac{a^2}{2}\left(\omega_s - \frac{\omega_0}{a}\right)^2\right) \exp(i\omega_s b) \quad (5)$$

Because the wavelet transform coefficient is the correlation one, the higher coefficient value is, the more similar the given signal to the wavelet with a certain scale. Thus the maximal scale value during shifting the wavelet through the fringe pattern is the fringe period and the coefficient value is the contrast of the fringe pattern at this position.

Now let the scale value with which the wavelet transform coefficient is maximal be $a_r$. Then

$$\omega_s = \frac{\omega_0}{a_r}, \tag{6}$$

Fig.1 shows a method of determining the local contrast of fringe pattern by CWT. Fig 1(a) shows the simulated fringe pattern with a quadric phase and hence a linearly increasing frequency, but different contrast from place to place. Fig. 1(b) shows local fringe contrast distribution obtained by wavelet transform. Because the wavelet function is the complex one, the wavelet coefficients are also complex and therefore their moduli are the fringe contrasts. We changed the scale parameter $a$ of the wavelet transform from 1 to 50 pixel and shifted the position parameter $b$ one by one pixel to get the correlation coefficient between the wavelet function and the given signal.

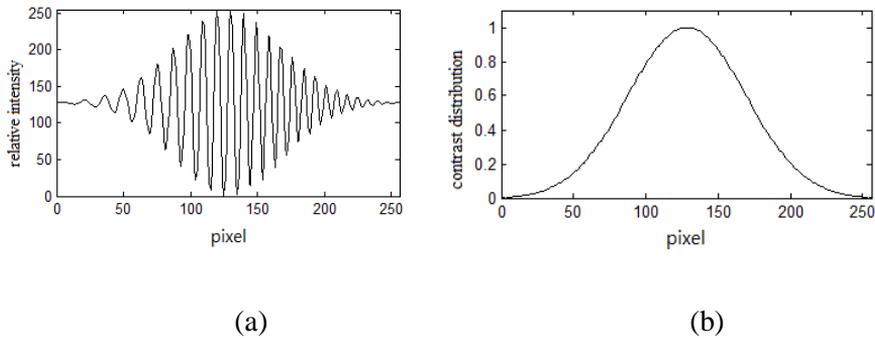

(a)          (b)

Fig. 1. Fringe relative intensity (a) and contrast distribution (b)

Fig. 2 shows the local contrast distribution of the 2-D interference fringe obtained by the above-mentioned method. Fig. 2(a) shows the computer generated 2-D fringe and Fig. 2 (b) shows the local contrast distribution of it.

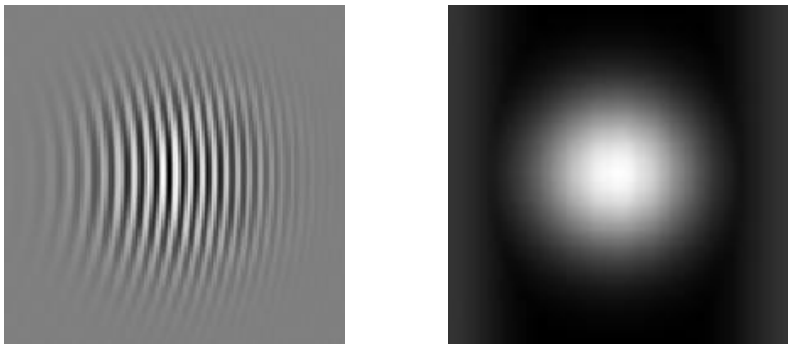

(a) (b)

Fig. 2 2-D interference fringe (a) and local contrast distribution (b)

## 3. Conclusions

In this paper we have proposed a method of determining the local contrast of interference fringe patterns using continuous wavelet transform. The CWT transforms the interference fringe pattern into the space-frequency domain.

The maximal scale value during shifting the wavelet through the fringe pattern is the fringe period and the coefficient value is the contrast of the fringe pattern at this position.